\documentclass[a4]{article}
\usepackage{graphicx}
\newcommand{\be}{\begin{equation}}
\newcommand{\ee}{\end{equation}}
\renewcommand{\k}[2]{\frac{#1}{#2}}

\newcommand{\pcd}[2]{\frac{d^n \! #1}{d #2^n}}

\def\s{\,\,\,\,\,}
\def\t{\tau}
\def\d{\delta}
\def\q{\Gamma_{d}}
\def\b{\beta}
\def\a{\alpha}
\def\z{\zeta}
\def\vv{\lambda}
\def\ra{\rightarrow}

\begin{document}
\title{\Large \bf Chaos in Small-World Networks}

\author{ Xin-She Yang \\
Department of Applied Mathematics and Department of Fuel and Energy, \\
        University of Leeds, LEEDS LS2 9JT, UK }

\date{ }
\maketitle

\begin{abstract}

A nonlinear small-world network model has been presented
to investigate the effect of nonlinear interaction
and time delay on the dynamic properties of small-world networks.
Both numerical simulations and analytical analysis for networks
with time delay and nonlinear interaction show chaotic features
in the system response when nonlinear interaction is strong enough
or the length scale is large enough. In addition, the small-world system
may behave very differently on different scales. Time-delay parameter
also has a very strong effect on properties such as the critical
length and response time of   small-world networks.\\

\end{abstract}


\noindent {\bf Citation detail:} X. S. Yang, Chaos in small-world networks,
{\it Phys. Rev. E} {\bf 63}, 046206 (2001).


\section{Introduction}

Since the pioneer work of Watts and Strogatz [1] on small-world networks,
there arise a lot of interesting research on the theory and application
of small-world networks [2-7]. The properties of complicated
networks such as internet servers, power grids, forest fires
and disordered porous media are mainly determined by the way
of connections  between the vertices or occupied sites.
One limiting case is the regular network with a high degree
of local clustering and a large average distance, while the other
limiting case is the random network with negligible local
clustering and small average distance. The small-world network
is a special class of networks with a high degree
of local clustering as well as  a small average distance.
Such small-world phenomenon can be obtained by adding randomly
only a small fraction  of the connections, and some common
networks such as power grids, film stars networks and
neural networks behaves like small-world networks [2-9].

The dynamic features such as spreading and reponse of the
network have also been investigated in recent studies [2,3]
by using shortest paths in system with sparse long-range
connections in the frame work of small-world models.
A simple time-stepping rule has been used to similuate the
the spreading of some influence such as a forest fire,
an infectious disease or a particle in percolating
media.  The influence propagates from the infected site
to all uninfected sites connected to it via a link at each
time step,  whenever a long-range connection or shortcut
is met, the influence is newly activated at the other end of
the shortcut so as to simulate long-range sparkling effect
such as the infect site (e.g., a person with influenza)
suddenly travels to a new place, or a portable computer
with virus that start to connect to the network a new site.
These phenomena have been sucessfully studied by Newman
and Watts model [2] and Moukarzel [3]. Their models are
linear model in the sense that the governing equation
is linear and the response is immediate as there is no
time delay in their models.

However, in reality, a spark or an infection can not start a
new fire spot or new infection immediately, it usually takes
some time $\Delta$, called ignition time or waiting time, to start
a new fire or infection. In addition, a fraction of infected site
shall recover after a futher time of $T$ to normality.
Thus the existing models are no longer be able  to predict
the response in the networks or systems with time delay.
Furthermore, the nonlinear effect such as the competition
factor as in the population dynamics, congestion feature such as
the traffic jam in internet communication and road networks,
and the frictional or viscous effect in the interaction of the
vertices. When considering these nonlinear effects, the resulting
small-world network model is generally no longer linear. Therefore,
a nonlinear model is yet to be formulated.

The main aim of this paper to present a more general nonlinear model
for the small-world networks by extending the existing Newman-Watts[2]
and Moukarzel [3] models to investigate the effects of
time-delay, site recovery and the nonlinear interaction due to
competition and congestion. The new model will
generally lead to a nonlinear difference differential
equation, whose solution is usually very difficult to
obtain if it is not impossible. Thus the numerical
simulation becomes essential [1]. However, we will take
the analytical analysis as far as possible and compare
with the results from numerical simulations. The characteristic
chaos of the network dynamics is then studied by reducing the
governing equation into a logistic equation. The control of
the chaos is also investigated by introducing the negative
feedback with time delay to the small-world networks.

\section{Nonlinear Model for Small-World Networks}

To investigate the nonlinear effect of time delay on
the properties of a small-world network, we now consider
a randomly connected network on a $d$-dimensional
lattice [1,2] (with $d=1,2,...$), and overlapping on
the network are a number of long-range
shortcuts randomly connecting some vertices,
and the fraction of the long-range shortcuts
or probability $p$ is relative small $p \ll 1$.
Now assuming an influence or a pollutant
particle spreads with a constant velocity $u=1$
in all directions and a newly infected site in the
other end of a shortcut will start but with a time
delay $\Delta$. Following the method by Newman and watts[2]
and Moukarzel [3], the total influenced volume
$V(t)$ comes from three contributions: one is the
influenced volume with $\q \int^{t}_0 t^{d-1} dt $
where $t$ is time and $\q$ is a shape factor,
the other contribution is $2 p V(t-\z-\Delta)$
for a hypersphere started at time $\z$. These two
components have used studied earlier [2,3] although
without the time delay parameter.  Now we add the
third component due to nonlinear interaction such
as friction, slow down due to congestion as in the
case of internet network and road traffic jam and lack of
other resource as lack of oxygen for the fire spark
to start a new fire.
By assuming this nonlinear effect as $-\mu V^2(t-\z-\Delta)$ where
$\mu \ll 1$ is a measure of nonlinear interaction.
By using a continuum approach to the network, then
$V(t)$ satisfies the following equation with time delay
\be
V(t)=\q \int^{t}_0 \z^{d-1}
[1+\xi^{-d} V(t-\z-\Delta)-\mu V^2(t-\z-\Delta)] d\z,
\label{equ-1}
\ee
where $d=1,2,...$ and $\q$ is shape factor of a hypersphere
in $d$-dimensions.  The {\it Newman-Watts} length scale
[2] can be conveniently defined as
\be
\xi=\k{2}{(p k d)^{1/d}},
\ee
where $k=const$ being some fixed range. By proper rescaling $t$
\be
\t=t (\q \xi^{-d} (d-1)!)^{1/d}, \s \d=\Delta (\q \xi^{-d} (d-1)!)^{1/d}.
\ee
and rewriting (\ref{equ-1}) in the rescaled form
\be
V(t)=\k{\xi^{d}}{(d-1)!} \int^{\t}_0 (\t-\z)^{d-1} [1+\xi^{-d} V(\z-\d)
-\mu V^2(\z-\d) ] d\z,
\label{equ-2}
\ee
we have a time-delay equation, after differentiating the
equation $d$ times
\be
\pcd{V}{\t}=\xi^{d}+ V(\t-\d) -\mu \xi^{d} V^2(\t-\d),
\label{equ-4}
\ee
which is a nonlinear delay differential equation, whose explicit
solutions is not always possible. In addition, the nonlinear
term and time delay can have strong effect on the behaviour
of the dynamic properties of the small-world networks.

\section{Chaos in Small-World Networks}

From the theory of dynamical systems, it is expected that
the dynamic features can be shown more clearly by using
the representation in Poincare plane [10], which usually transform
a nonlinear differential equation into a nonlinear iterated
map or logistic equation. Now we write equation (\ref{equ-4})
in a difference form and take $d \t=\d$ to get a logistic
equation. In order to focus on the main characteristics of
the dynamics, for simplicity, we can take $\d=1$ in 1-D (or $d=1$),
and we then have
\be
V_{n+1}=\xi + 2 V_n-\mu \xi V_n^2,        \label{equ-5}
\ee
where $V_{n+1}=V(\t)$ and $V_n=V(\t-1)$. By changing variables
\be
v_{n+1}=\k{(2+2A\mu \xi)}{\mu \xi}(V_{n+1}+A), \s v_{n}=\k{(2+2A\mu \xi)}{\mu \xi}(V_{n}+A), \s A=\k{\sqrt{1+4 \mu \xi^2}-1}{2 \mu \xi},
\label{equ-6}
\ee
we can rewrite equation (\ref{equ-5}) as
\be
v_{n+1}=\vv v_{n} (1-v_{n}), \s \vv=(\sqrt{1+4 \mu \xi^2}+1),
\ee
which is a standard form of well-known logistic equation [10].
Since the parameter range of $\vv$ for period doubling and
chaos is known, we have
\be
\mu \xi^2=\k{(\vv-1)^2-1}{4}.
\ee
The system becomes chaotic as $\vv$ is bigger than $\vv_* \approx 3.5699$
but usually below 4.0, so the chaos begin at
\be
\xi_*=\sqrt{\k{1.401}{\mu}}, 
\label{equ-77}
\ee
For $\vv$ less than $\vv_0 \approx 3.0$, the system approach to a fixed point,
that is
\be
\xi_0=\sqrt{\k{0.75}{\mu}}, 
\ee
For a fixed $\mu$, when $\xi_0 < \xi < \xi_*$, then $\vv < \vv_*$,
the system is in a period doubling cascade.
When $\xi > \xi_*$, the system is chaotic. Clearly,
as $\mu \ra 0$, $\xi_* \ra
\infty$. The system behavior depends on the lengthscale of small-world
networks. The system may looks like chaotic on a large scale
greater than the critical length scale $\xi_*$ and the same system
may be well regular on the even smaller scale. So the system behaves
differently on different scales.

To check the analytical results, we also simulated the scenario by using
the numeric method [1,2] for a network size $N=500,000$, $p=0.002$
and $k=2$ on a 1-D lattice. Different values of the nonlinear interaction
coefficient $\mu$ are used and the related  critical length $\xi_*$
when the system of small-world networks becomes chaotic. Figure 1 shows
$\xi_*$ for different values of $\mu$. The solid curve is the analytical
results (\ref{equ-77}) and the points (marked with $\circ$) are numerical
simulations. The good agreement verifies the analysis. However, as the
typical length increases, the difference between these two curves becomes
larger because the governing equation is main for infinite size network.
So the difference is expected due to the finite size of the network
used in the simulations.

\section{Feedback and Chaos Control of Small-World Networks}

The occurrance of the chaos feature in small-world networks is due to the
nonlinear interaction term and time delay. This chaotic
feature can be controlled by adding a negative feedback term [11,12]. In
reality, the influence such
as a signal or an influence (e.g.. influenza) only
last a certain period of time $T$, then
some of the influenced sites recover to normality. From the small-world
model equation (\ref{equ-1}), we see that this add an extra term
$\b V (t-\Delta-T)$, which means that a fraction ($\b$) of the infected
sites at a much earlier time ($t-\Delta-T$) shall recover at $t$.
So that we have the modified form of equation
(\ref{equ-4}) as
\be
\pcd{V}{\t}=\xi^{d}+ V(\t-\d) -\mu \xi^{d} V^2(\t-\d)-\b \xi^d V(\t-\d-\t_0),
\label{equ-7}
\ee
where $\t_{0} =T (\q \xi^{-d} (d-1)!)^{1/d}$. For $d=1$, we can take
$\t_0=j \d$ (j=1,2,...) without losing its physical importance. By using the transform (\ref{equ-6}), we have the modified form of the logistic equation
\be
v_{n+1}=\vv v_n (1-v_n)+ \a (v_n-v_{n-j}), \s j=1,2,...
\label{equ-8}
\ee
with
\be
\vv=(\sqrt{1+4 \mu \xi^2}+1), \s \a=\b \xi,
\ee
which is essential the Escalona and Parmananda form [12] of OGY
algorithm [11] in the chaos control strategy. We can also write
(\ref{equ-8}) as
\be
v_{n+1}=\Lambda v_n (1-v_n)-\a v_{n-j}, \s j=1,2,...
\label{equ-10}
\ee
with
\be
\Lambda=(\sqrt{(1-\b \xi)^2+4 \mu \xi^2}+1), \s \a=\b \xi.
\ee
This last form (\ref{equ-10}) emphases the importance of time delay
and the effect of negative feedback in controlling the chaos.

For a fixed value of $\Lambda=3.8$, we find a critical
vaule of $\a_*=0.27$ for $j=1$ and $\a_*=0.86$ for $j=2$
to just control the chaos so that the system settles to a fixed
point. For $\a < \a_*$, the feedback is
not strong enough and the chaos is not substantially suppressed.
For $\a > \a_*$, the strong feedback essentially control the chaos
of the small-world networks. Figure 2 shows that the effect of
recovery of the infected site or the delay feedback on the system
behavior. The dotted points are for the chaotic response when there
is no feedback  ($\Lambda=3.8, \,\a=0$), while the solid curve
corresponds to the just control of the chaos by feedback ($\Lambda=3.8,
\, \a=0.27$). This clearly indicates that the proper feedback
due to healthy recovery and time delay can control the chaotic
reponse to a stable state.

\section{Discussion}

A nonlinear small-world network model has been presented here to
characterise the effect of nonlinear interaction, time delay,
and recovery on small-world networks. Numerical simulations and
analytical analysis for networks with time delay and nonlinear
interaction show that the system reponse of the small-world networks
may become chaotic on the scale greater than the critical
length scale $\xi_*$, and at the same time the system may still
quite regular on the smaller scale. So the small-world system
behaves differently on different scales. Time-delay parameter $\d$
has a very strong effect on properties such as the critical
length and reponse time of the networks.

On the other hand, in order to control the possible chaotic behaviour
of small-world network, a proper feedback or healthy recovery of
the infected sites is needed to stable the system reponse. For a
negative delay feedback, comparison of different numerical simulations
suggests that a linear recovery rate $\b$ or a linear feedback
can properly control the chaos if the feedback is strong
enough. This may has important applications in the management and
control of the dynamic behaviour of the small-world networks.
This shall be the motivation of some further studies of the dynamics
of small-world networks.

\begin{figure}

\centerline{ \includegraphics[width=4in,angle=270]{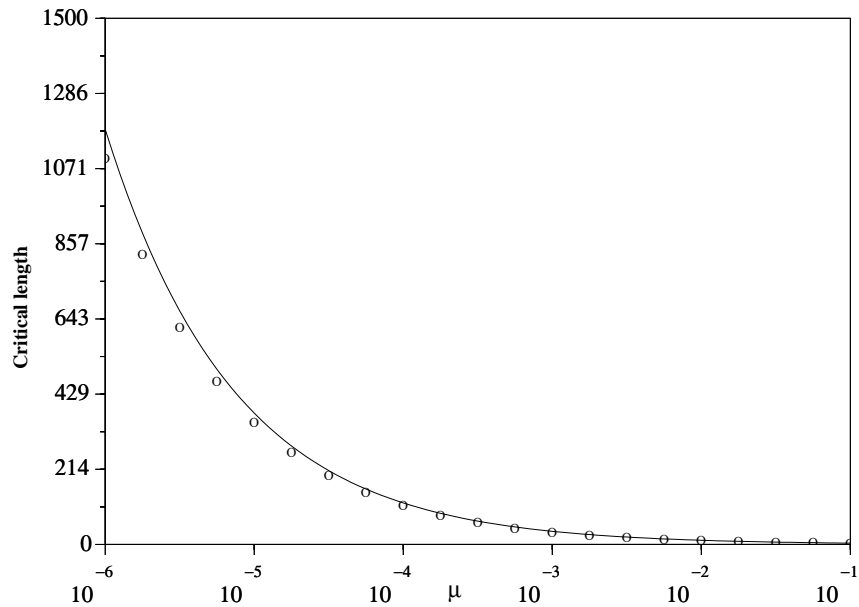} }

\caption{ Critical length versus the nonlinear
interaction coefficient $\mu$  for a network size $N=500,000$,
and $p=0.002$. Numerical results (marked with $\circ$)
agree well with analytical express (solid). }
\end{figure}

\begin{figure}

\centerline{ \includegraphics[width=4in,angle=270]{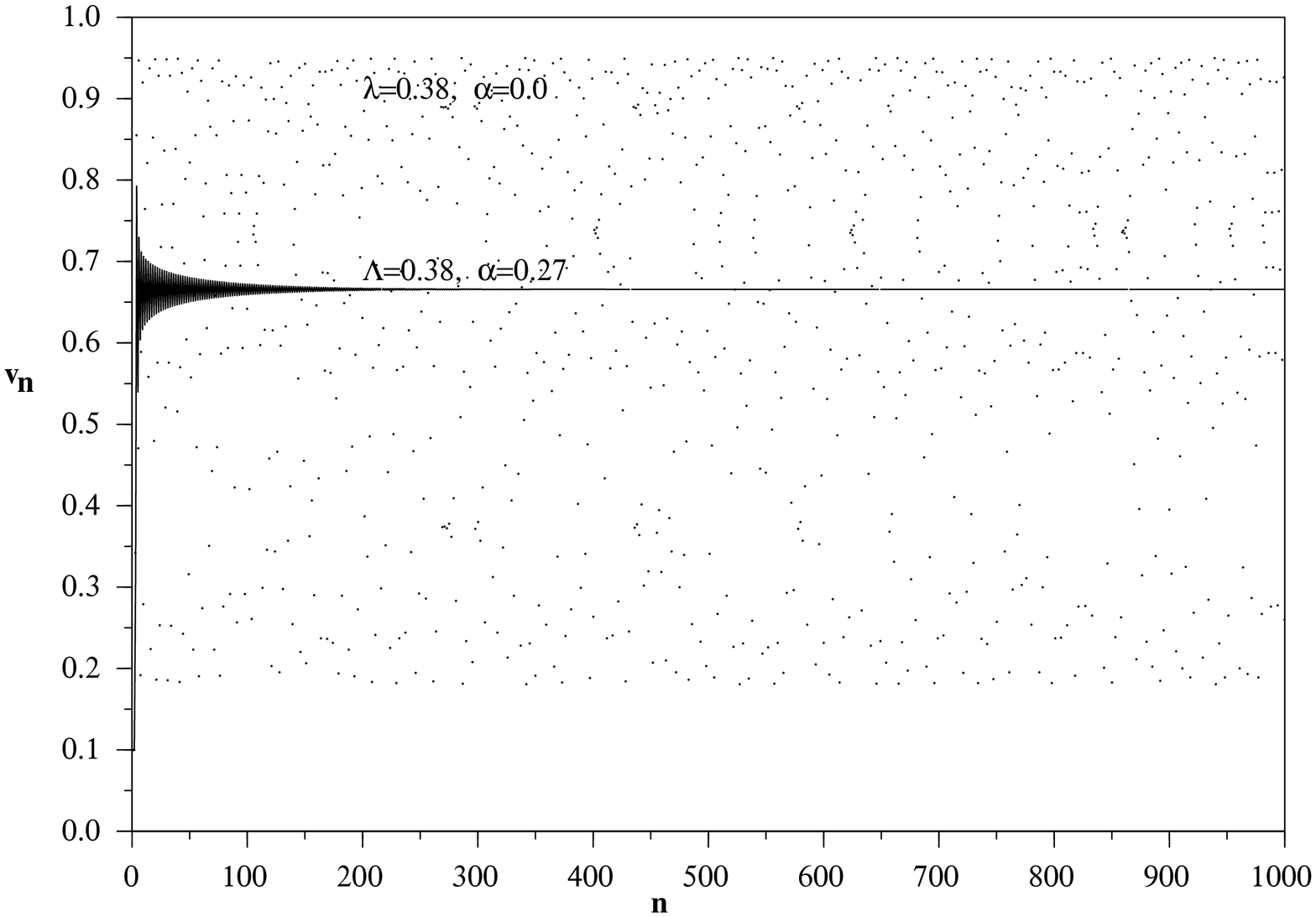} }

\noindent Figure 2 Comparison of chaos and chaos control due to
delay feedback.   The dotted points are for the chaotic response when there
is no feedback  ($\Lambda=3.8, \,\a=0$), while the solid curve
corresponds to the just control of the chaos by feedback ($\Lambda=3.8,
\, \a=0.27$).
\end{figure}

\end{document}